\documentclass[aps,prd,showkeys,nofootinbib,amsmath,amssymb,superscriptaddress,onecolumn]{revtex4-2}
\usepackage{booktabs}
\usepackage[dvipsnames]{xcolor}
\usepackage{dsfont}
\usepackage{mathtools} 
\definecolor{nicegreen}{rgb}{0.07, 0.564, 0.04}
\usepackage{framed}
\usepackage{comment}
\usepackage{xspace}
\usepackage{graphicx}
\usepackage{dcolumn}
\usepackage{bm}
\usepackage{tikz-cd}
\usepackage[normalem]{ulem}
\usepackage{slashed}
\usepackage{accents}
\usepackage{forest}
\usepackage{soul}

\newcommand\ie{\mbox{\textit{i.\,e.}}\xspace}
\newcommand\eg{\mbox{e.\,g.}\xspace}
\newcommand\hatslashed[1]{{\hat{#1}\mathllap{\slashed{#1}}}}

\newcommand\D{\mathrm{d}}

\newcommand{\hp}{\hat{p}}
\newcommand{\hk}{\hat{k}}
\newcommand{\hpi}{\hat{\pi}}

\newcommand{\hH}{\hat{H}}
\newcommand{\mompar}{\dot{\partial}}
\newcommand{\hPi}{\hat{\Pi}}
\newcommand{\id}{\mathds{1}}

\newcommand{\Fabcom}[1]{\textcolor{orange}{\bf [#1]}}

\newcommand{\ord}[2]{\accentset{(#2)}{#1}}
\newcommand{\SProd}{\star}

\DeclarePairedDelimiter\braket{\langle}{\rangle}
\DeclarePairedDelimiterX\Braket[2]{\langle}{\rangle}{#1 \delimsize\vert #2}
\newcommand{\hbpI}{\hat{\bar{\pi}}^{(1)}}

\begin{document}

\title{Spin couplings as witnesses of Planck scale phenomenology}

\author{Pasquale Bosso}\email[]{pbosso@unisa.it}
\affiliation{Dipartimento di Ingegneria Industriale, Universit\`a degli Studi di Salerno, Via Giovanni Paolo II, 132 I-84084 Fisciano (SA), Italy}
\affiliation{INFN, Sezione di Napoli, Gruppo collegato di Salerno, Via Giovanni Paolo II, 132 I-84084 Fisciano (SA), Italy}

\author{Fabrizio Illuminati}\email[]{filluminati@unisa.it}
\affiliation{Dipartimento di Ingegneria Industriale, Universit\`a degli Studi di Salerno, Via Giovanni Paolo II, 132 I-84084 Fisciano (SA), Italy}
\affiliation{INFN, Sezione di Napoli, Gruppo collegato di Salerno, Via Giovanni Paolo II, 132 I-84084 Fisciano (SA), Italy}

\author{Luciano Petruzziello}\email[]{lupetruzziello@unisa.it}
\affiliation{Dipartimento di Ingegneria Industriale, Universit\`a degli Studi di Salerno, Via Giovanni Paolo II, 132 I-84084 Fisciano (SA), Italy}
\affiliation{INFN, Sezione di Napoli, Gruppo collegato di Salerno, Via Giovanni Paolo II, 132 I-84084 Fisciano (SA), Italy}
\affiliation{Institut f\"ur Theoretische Physik, Albert-Einstein-Allee 11, Universit\"at Ulm, 89069 Ulm, Germany}

\author{Fabian Wagner}
\email[]{f.wagner@thphys.uni-heidelberg.de}
\affiliation{Dipartimento di Ingegneria Industriale, Universit\`a degli Studi di Salerno, Via Giovanni Paolo II, 132 I-84084 Fisciano (SA), Italy}
\affiliation{INFN, Sezione di Napoli, Gruppo collegato di Salerno, Via Giovanni Paolo II, 132 I-84084 Fisciano (SA), Italy}
\affiliation{Institut f\"ur theoretische Physik, Universit\"at Heidelberg, Philosophenweg 16, 69120
Heidelberg, Germany}

\date{\today}

\begin{abstract}
Modified dispersion relations (MDRs) and noncommutative geometries are phenomenological models of Planck-scale corrections to relativistic kinematics, motivated by several approaches to quantum gravity. High-energy astrophysical observations, while commonly used to test such effects, are limited by significant systematic uncertainties. In contrast, low-energy, nonrelativistic experiments provide greater control, with precision serving as an amplifier for Planck-suppressed corrections. We derive corrections to Pauli’s equation for nonrelativistic spin-$1/2$ particles in a magnetic field, incorporating general MDRs and noncommutative geometries. Applying our framework to $\kappa$-Poincar\'e symmetries and minimal-length quantum mechanics, we identify Planck-scale corrections accessible in the nonrelativistic regime. Using the electron’s anomalous magnetic moment, we constrain model parameters, pushing the $\kappa$-Poincar\'e scale in the bi-crossproduct representation beyond $10^{10} \mathrm{GeV}.$ These results highlight the complementarity of low-energy precision tests and astrophysical observations in probing quantum gravity phenomenology.
\end{abstract}

\maketitle

\section{Introduction}

The unification of quantum mechanics and general relativity remains a central open problem in theoretical physics. Together, they form the effective framework of semiclassical gravity, which is highly successful as an effective theory, but breaks down in extreme environments such as black hole interiors or the early universe. Attempts to quantize gravity perturbatively, analogous to standard quantum field theories, fail at the Planck scale \cite{Goroff:1985sz}, and thus require UV completion. Despite numerous proposals for UV completions \cite{Hawking:1975vcx,Weinberg79,Witten:1981me,Rovelli:1989za,Reuter:1996cp,Ashtekar:2004vs}, no consensus has emerged on a definitive theory.

When consistency arguments fail to single out a preferred model, experimental input becomes essential. For many years, the inaccessibility of the Planck scale ($m_{\rm pl} \sim 10^{19}\mathrm{GeV}$) seemed to preclude such input. However, increases in control over mesoscopic quantum systems \cite{Bose:2017nin,Marletto:2017kzi,Donadi:2020kzc} as well as precision in astrophysical observations \cite{Amelino-Camelia:1999hpv,mdrs1,HESS:2011aa,IceCube:2018dnn,MAGIC:2020egb,LHAASO:2024lub} have led to the emergence of quantum gravity phenomenology \cite{Amelino-Camelia:1997ieq} in the past two decades (see \cite{mdrs2,Addazi:2021xuf} for recent reviews). This program aims to identify potential quantum-gravity effects that can be amplified sufficiently to produce measurable corrections at accessible energy scales. 

Several approaches to quantum gravity, including string theory \cite{Kostelecky:1988zi,Magueijo:2004vv,Mavromatos:2010pk}, loop quantum gravity \cite{Girelli:2012ju,Amelino-Camelia:2016gfx}, asymptotically safe quantum gravity \cite{Girelli:2006sc,Calmet:2010tx}, causal dynamical triangulations \cite{Coumbe:2015aev}, and three-dimensional quantum gravity \cite{Matschull:1997du,Freidel:2003sp,Freidel:2005me,Freidel:2005ec,Freidel:2005bb,Meusburger:2008dc,Rosati:2017toi}, suggest deviations from the standard relativistic energy-momentum relation. These modified dispersion relations (MDRs) are expected to become significant near the Planck scale and serve as a potential window into quantum gravity. MDRs offer insights into high-energy physics beyond current particle accelerators, with phenomenological consequences such as energy-dependent variations in the speed of light \cite{Amelino-Camelia:1997ieq,Magueijo:2000zt,Magueijo:2001cr} or potential violations of the equivalence principle \cite{Magueijo:2001cr,Magueijo:2002xx,Wagner:2023fmb,Hohmann:2024lys}.

Modified dispersion relations are not Lorentz invariant. This may reflect spontaneous violation of Lorentz invariance (LIV) \cite{Kostelecky:1988zi}, as studied in the framework of the standard model extension \cite{Kostelecky:2008ts}. Alternatively, spacetime-symmetry transformations may be deformed to retain a relativity principle, as in deformed or doubly special relativity (DSR) \cite{Magueijo:2000zt,Amelino-Camelia:2000stu,Amelino-Camelia:2002cqb}, see \cite{Amelino-Camelia:2010lsq} for a review. A prominent deformation of spacetime symmetries is the $\kappa$-Poincar\'e algebra \cite{Majid:1994cy,Kowalski-Glikman:2002iba}.

Models of deformed spacetime symmetries are often formulated on noncommutative geometries, where coordinates obey nontrivial commutation relations. For instance, the $\kappa$-Poincar\'e algebra represents a Lie-algebra type of noncommutativity dubbed $\kappa$-Minkowski spacetime, for recent reviews see \cite{Arzano:2021scz,Kowalski-Glikman:2022xtr}. Field theories on such geometries are constructed through deformation quantization \cite{Moyal:1949sk,Kontsevich:1997vb}, where fields are defined on ordinary spacetime but equipped with a nontrivial $*$-product, see \cite{Douglas:2001ba,Hersent:2022gry,Moshayedi:2022dkl} for reviews and \cite{Vitale:2023znb} for a recent introduction.

While high-energy experiments seem natural for probing Planck-scale effects, low-energy experiments (\ie experiments in the nonrelativistic regime) also provide competitive constraints \cite{Amelino-Camelia:2009wvc,Wagner:2023fmb,Hohmann:2024lys}. Their higher precision acts as a substitute for energy in amplifying small effects. For example, the anomalous magnetic moment of the electron has been determined up to 13 digits of precision. Besides, astrophysical tests of LIV and DSR models, while valuable, often face significant systematic uncertainties due to incomplete knowledge of the underlying sources. In contrast, table-top experiments offer much greater control over measured quantities. Thus, the nonrelativistic regime provides a complementary testing ground to astrophysical observations.

Considering nonrelativistic spinors, rather than just scalars \cite{Wagner:2023fmb}, is essential due to their relevance in high-precision table-top experiments. Measurements, for example, of the anomalous magnetic moment of the electron, or the CHSH inequality are commonly performed on nonrelativistic spin-$1/2$ particles, namely electrons. These experiments offer a highly controlled environment, making spin-$1/2$ systems an effective probe of Planck-scale modifications. 

In this work, we derive the modifications to Pauli's equation, which governs the dynamics of nonrelativistic spin-$1/2$ particles in a background magnetic field arising from a general MDR and a general noncommutative geometry. Thus, our approach captures aspects of both symmetry breaking and symmetry deformation. We identify Planck-scale corrections that are most accessible to nonrelativistic experiments and constrain model parameters using existing data on the anomalous magnetic moment of the electron. 

In particular, starting from the Dirac equation for a spinor charged under $U(1)$, we incorporate the MDR through a modified Dirac operator and replace the standard product of fields with a general $*$-product. The nonrelativistic limit of this modified Dirac equation then yields the Pauli equation. Note that complementary results for a specific model with different modifications inspired by loop quantum gravity have been recently obtained in \cite{Melo:2024gxl}.

We apply our approach to four models: the $\kappa$-Poincar\'e algebra in the bi-crossproduct and classical representations, as well as two widely studied minimal-length models \cite{Kempf:1994su}, also known as commutative spatial Snyder geometries.\footnote{See \cite{Hossenfelder:2012jw} for a review and \cite{Bosso:2023aht} for a critical assessment of the field.} Consistent with earlier results for scalar fields \cite{Wagner:2023fmb}, we find that the bi-crossproduct representation is significantly better suited for nonrelativistic experiments than minimal-length models, which are often considered in the IR. Finally, we use the electron's magnetic moment to place constraints on the model parameters.

The paper is organized as follows: Sections \ref{sec:MDR} and \ref{sec:NonCom} provide an overview of MDRs and noncommutative geometry, respectively. In Section \ref{sec:spin_coupling}, we introduce the general deformed Dirac equation and derive its nonrelativistic limit, first without external field, and then with $U(1)$ gauge coupling. Section \ref{sec:examples} illustrates these results using the $\kappa$-Poincar\'e model and minimal-length quantum mechanics. Section \ref{sec:anomalous_magmom} explores the impact of the modified spin coupling on the electron’s magnetic moment, compares theoretical predictions to experimental data, and constrains the modification parameters. Finally, in Section \ref{sec:conclusions} we discuss our results and present an outlook on possible future developments.

Throughout the paper, spacetime (spatial) indices are indicated by Greek (Latin) letters, we raise and lower indices with the Minkowski metric $\eta_{\mu\nu}$ with mostly positive signature and we work with the units in which $\hbar=1$. 

\section{Modified dispersion relations}
\label{sec:MDR}

For phenomenological purposes, it is not necessary to specify an MDR to all orders in the Planck scale. Assuming $\ell$ represents the scale at which MDR effects become significant\footnote{In standard quantum-gravity considerations, this corresponds to the Planck length, $\ell = l_{\text{P}} \sim 10^{-35} , \mathrm{m}$.}, we consider a set of MDRs expanded to linear order in $\ell$. The most general analytic form of such MDRs is given by \cite{Wagner:2023fmb}:
\begin{equation}
    M^2c^4=-k^\mu k_\mu c^2+\ell\sum_{n=0}^2\sum_{m=0}^{3-n} a_{n,m}(Mc)^n k^m \left(\frac{E}{c}\right)^{3-n-m}\equiv\mathcal{C}^{(0)}+\ell\mathcal{C}^{(1)},\label{eqn:mdr}
\end{equation}
with dimensionless model parameters $a_{n,m}$ and the spatial momentum along some slicing $k_i$, and where $k^2=\delta^{ij}k_i k_j.$

The last equality defines the Poincar\'e mass Casimir $\hat{\mathcal{C}}^{(0)}=\hk^\mu\hk_\mu c^2$ and its correction $\hat{\mathcal{C}}^{(1)}.$
Corrections proportional to the parameter $a_{0,3}$ ($\sim\ell k^3$) are predicted by Lorentz-violating theories like the standard model extension \cite{Kostelecky:2008ts}, conventional models of minimal-length quantum mechanics \cite{Wagner:2023fmb} and a comparably little known basis of deformed special relativity \cite{Girelli:2006fw,Amelino-Camelia:2014rga}.
Furthermore, corrections going with the parameter $a_{0,2}$ ($\sim\ell E k^2$) are predicted by the bi-crossproduct representation of DSR \cite{Kowalski-Glikman:2002iba}. 

The appearance of spatial momenta and energies in Eq. \eqref{eqn:mdr} indicates the presence of a unit, timelike, dimensionless deformation vector $d^\mu.$ Thus, in {covariant notation}, Eq. \eqref{eqn:mdr} reads
\begin{equation}
    M^2c^4=-k^\mu k_\mu c^2+\ell\sum_{n=0}^2\sum_{m=0}^{3-n} a_{n,m}(Mc)^n k_\perp^m k_d^{3-n-m},\label{eqn:mdrkdkperp}
\end{equation}
where we introduced the scalars
\begin{align}
    k_d &= d^\mu k_\mu ,\\
    k_\perp^2 &=k^\mu k_\mu+ k_d^2 \simeq -M^2c^2+k_d^2.\label{eqn:kperp}
\end{align}
Here, the last equality holds up to the order $\mathcal{O}(\ell^0)$ and on the mass shell. Thus, when applying the on-shell relation, we can effectively rewrite Eq. \eqref{eqn:mdr} as
\begin{equation}
    M^2c^4\simeq -k^\mu k_\mu+\ell\sum_{n=0}^2\sum_{m=0}^{3-n} a_{n,m}(Mc)^n k_\perp^{m}(M^2c^2+k_\perp^2)^{\frac{3-n-m}{2}}.\label{eqn:mdrkd}
\end{equation}
{In this paper, we examine the effective nonrelativistic dynamics implied by this type of MDR. Beyond MDRs, noncommutative geometries are another common feature of Planck-scale deformed models, which we address in the next section.}

\section{Noncommutative geometry}
\label{sec:NonCom}

{
On a noncommutative geometry, even at the classical level, coordinates are operators acting on a Hilbert space $\mathcal{H}$. Thus, they satisfy a nontrivial commutation relation
\begin{equation}
    [\hat x^\mu,\hat x^\nu]=i\Theta^{\mu\nu} (\hat{x},\hat p),\label{eqn:CoordNonCom}
\end{equation}
with the noncommutativity tensor $\Theta^{\mu\nu}$ which has units of squared length.  
As the coordinates become operators so do classical fields $\psi(\hat{x})$. The operator property can be modeled by working with auxiliary fields $\psi_c(x)$ defined on a commutative manifold (on which coordinates $x$ are not operators), while replacing the ordinary product among them with a (possibly noncommutative, but associative) star product $\SProd.$ This star product has to be chosen such that
\begin{equation}
    [x^\mu,x^\nu]_\SProd=x^\mu\SProd x^\nu-x^\nu\SProd x^\mu=i\Theta^{\mu\nu}\label{eq:SPCond}
\end{equation}
In the commutative limit the star product reduces to the standard product:
\begin{align}
    \lim_{\ell\to0}\psi(x)_c\SProd\tilde\psi_c(x)=\psi_c\tilde\psi_c.
\end{align}
Beyond the $\SProd$-product, noncommutative geometries require an involution ${}^\dagger$ that reduces to Hermitian conjugation ${}^*$ in the commutative limit (see \cite{Hersent:2022gry,Vitale:2023znb} for more details.)

When $\Theta^{\mu\nu}=\Theta^{\mu\nu}(\hat x),$ leading-order corrections can be modeled as 
\begin{align}
    \psi(x)_c\SProd\tilde\psi_c(x)=\psi_c\tilde\psi_c +\frac{i}{2}\bar\Theta^{\mu\nu}\partial_\mu \psi_c\partial_\nu \tilde\psi_c,\label{eqn:KontsevichProd}
\end{align}
where the tensor structure of $\bar\Theta^{\mu\nu}$ has to be constructed from the tensors $x^\mu,$ $d^\mu$ and $\eta^{\mu\nu},$ and, by virtue of Eq. \eqref{eq:SPCond} has to satisfy the relation $\bar\Theta^{[\mu\nu]}=\Theta^{\mu\nu}.$ The exact $\star$ product depends on the chosen representation of the algebra given in Eq. \eqref{eqn:CoordNonCom}, and the realization, \ie the representation of the coordinates on the Hilbert space $\mathcal{H}$ \cite{Meljanac:2006ui,Meljanac:2007xb}. For example, the Kontsevich star product \cite{Kontsevich:1997vb} is such that $\bar\Theta^{\mu\nu}=\Theta^{\mu\nu}.$ 

For geometries for which $\Theta^{\mu\nu}$ additionally depends on momenta, such as Snyder models, the scalar product must be derived explicitly. To do so, it is necessary to choose a realization, and follow a procedure of the kind outlined in Refs. \cite{Battisti:2010sr,Meljanac:2016gbj,Meljanac:2020fde,Meljanac:2021jdr,Meljanac:2021qgq}. As we will not work with generalized noncommutative Snyder models in the following, we assume the approximation in Eq. \eqref{eqn:KontsevichProd} to be applicable. 

For this paper, we assume the noncommutative geometry to be modeled by a hermitian realization.
A hermitian realization is a representation in which noncommutative positions (as expressed in terms of commutative positions and momenta) are Hermitian with respect to the Lebesgue measure. Hermitian realizations ensure that, for Moyal space \cite{Szabo:2001kg}, $\kappa$-Minkowski spacetime in the classical basis \cite{Meljanac:2010ps}, and generalized Snyder spaces \cite{Meljanac:2017ikx}, the integral property
\begin{equation}
    \int\D^4x ~ \psi_c(x)^\dagger\SProd \tilde\psi_c(x)=\int\D^4x ~ \psi_c^*(x)\tilde\psi_c(x)\label{eqn:HermReal}
\end{equation}
is satisfied.
While this property has been conjectured to hold for Hermitian realizations in general \cite{Meljanac:2017ikx}, a formal proof remains outstanding.
Here, we follow this conjecture for reasons of simplicity. Note that Eq. \eqref{eqn:HermReal} only applies to products of two fields.
For field theories this implies that the star product is trivial for kinetic terms and mass terms, but becomes nontrivial for interactions. This is analogous to nonlocal Lorentzian field theories, for which the nonlocality can be shifted from the kinetic term to the interactions \cite{Buoninfante:2018mre}.

Having outlined the basics of MDRs and noncommutative geometries, we now proceed to discuss some of the subtleties of the nonrelativistic limit in deformed models.
}

\section{Nonrelativistic limit}
\label{sec:NRMDR}

The transition from the Lorentzian to the Galilean regime implies a number of changes to the underlying theory. First, it amounts to an expansion in velocities, which are to be small compared to the speed of light.
Second, it implies that general tensor fields $T_{\mu_1\dots\mu_n}^{\nu_1\dots\nu_m},$ irrespective of additional spin degrees of freedom, are to be expanded as
\begin{equation}
    T_{\mu_1\dots\mu_n}^{\nu_1\dots\nu_m} = \sum_{i=0}^{\infty}\frac{\ord{T}{i}_{\mu_1\dots\mu_n}^{\nu_1\dots\nu_m}}{c_i},
    \label{eqn:TenNot}
\end{equation}
such that the expansion elements $\ord{T}{i}_{\mu_1\dots\mu_n}^{\nu_1\dots\nu_m}$ are independent of the speed of light. As a change of coordinates can involve the speed of light, this expansion is clearly coordinate-dependent.

To identify an object as moving slowly, we need a reference system to measure the relative velocity to.
This necessitates introducing a Newtonian (global) notion of time, represented by a fixed, dimensionless timelike vector field $\tau^\mu$. For table-top experiments, this vector field is naturally provided by the four-velocity of Earth along its trajectory, \ie, the lab frame. This division into time and space establishes a preferred time coordinate $t$ satisfying
\begin{equation}
    \D t=\tau_\mu\D x^\mu.
\end{equation}
Supplementing time with spatial coordinates, we construct the adapted frame $(t,x_i)$.
It is in this coordinate frame that the $1/c$-expansion in Eq. \eqref{eqn:TenNot} is consistent, greatly simplifying the analysis.

Furthermore, by means of $\tau^\mu$, we can define the Hamiltonian as well as the spatial momenta
\begin{equation}
    H\equiv \tau^\mu k_\mu c=k_t c,\qquad k^2\equiv k^\mu k_\mu+(\tau^\mu k_\mu)^2=k_ik_i.
\end{equation}
These quantities do not necessarily coincide with the projections $k_d$ and $k_\perp$ because the vector fields $d^\mu$ and $\tau^\mu$ may differ from each other. {For instance, consider $\tau^\mu$ to be comoving with the Earth so that the nonrelativistic limit is taken in the lab frame. Then, considering the peculiar motion of the Earth relative to the Sun, the solar system relative to the galactic center and the galaxy relative to the whole Universe (say the cosmic microwave background), it is unlikely that $d^\mu$ and $\tau^\mu$ coincide, unless there is a symmetry linking both (like the $\kappa$-Poincar\'e symmetry).}
We denote the deviation between the two vectors as
\begin{equation}
    \mathcal{A}^\mu=\tau^\mu-d^\mu.
\end{equation}
As shown in \cite{Wagner:2023fmb}, timelike vectors must coincide to lowest order in $1/c$ to ensure a consistent nonrelativistic limit, \ie, because motion at this level is restricted to the direction of the global time.
However, deviations appear for the spatial components at order $1/c.$ In other words, in adapted coordinates
\begin{equation}
    \mathcal{A}_\mu=\frac{\delta_{i\mu}\ord{\mathcal{A}}{1}_i}{c}+\mathcal{O}(c^{-2}).
\end{equation}
The spatial vector $\ord{\mathcal{A}}{1}_i$ equals the relative speed the reference system possesses with respect to the deformation.

In LIV models, it is unlikely that the relative speed to the deformation is negligibly small on Earth, just because the Earth's motion is influenced by many factors. To estimate the order of magnitude of such velocity, we notice that {in LIV models} the deformation vector produces a preferred frame at the cosmological level.
Indeed, in cosmology there is one preferred frame, the rest frame of the CMB (Cosmic Microwave Background). Generally, the Earth's motion relative to this frame amounts to $10^{-3}c$. During the Earth's orbit around the sun, and the Sun's orbit around the galactic center, this relative velocity varies by about the same amount.
Thus, we can estimate
\begin{equation}
    |\ord{\mathcal{A}}{1}_i|\sim 10^{-3}c,\label{eqn:EstimatedDeformation}
\end{equation}
which in the context of nonrelativistic physics is indeed significant. {Note that this estimate does not apply to DSR models which retain a relativity principle.}

{Next, we apply the concepts reviewed in the three previous Sections to construct a deformed model of nonrelativistic spinor dynamics.}

\section{Deformed spin coupling}
\label{sec:spin_coupling}

Recent work \cite{Wagner:2023fmb} has demonstrated that experiments in the IR impose competitive constraints on some of the MDRs in Eq. \eqref{eqn:mdr}, while minimal-length models are particularly ill-suited for testing at low energies. However, that study neglected spin by focusing on the nonrelativistic limit of the Klein-Gordon equation and did not account for noncommutative geometries, thereby explicitly violating Lorentz invariance. To accurately describe ordinary matter, it is essential to consider the Dirac equation. In fact, in the context of minimal-length quantum mechanics, the nonrelativistic limit of the Dirac equation has been shown to introduce corrections to spin coupling \cite{Bosso:2022ogb}. Here we generalize this result to particles with spin and noncommutative geometries.

{We assume that the general structure of the Dirac Lagrangian on a general noncommutative geometry with MDR continues to be characterized by a Dirac operator $\hatslashed{\pi}.$ Here we introduced Dirac's slash notation $\hatslashed{\pi}=\gamma^\mu\hpi_\mu,$ with the Dirac matrices $\gamma^\mu$ satisfying the Clifford algebra
\begin{equation}
    \{\gamma^\mu,\gamma^\nu\}=-2\eta^{\mu\nu}.
\end{equation}
In the limit $\ell\to 0,$ the Dirac operator ought to recover its ordinary shape, \ie $\lim_{\ell\to 0}\hatslashed{\pi}=\hatslashed{k}.$ As a result, the Dirac Lagrangian reads\footnote{Hereafter, we drop the subscript $c$ for commutative fields for notational simplicity because all fields are defined on a commutative manifold.}
\begin{equation}
    \mathcal{L}=\psi^\dagger\gamma^0\SProd(\hatslashed{\pi}-Mc)\psi.\label{eqn:DiracLagNoCharge}
\end{equation}
Therefore, applying Eq. \eqref{eqn:HermReal} at the level of the action we obtain
\begin{equation}
    S=\int\D^4x\psi^\dagger\gamma^0\SProd(\hatslashed{\pi}-Mc)\psi=\int\D^4x\bar\psi(\hatslashed{\pi}-Mc)\psi,
\end{equation}
where $\bar\psi=\psi^*\gamma^0.$ As expected, for Hermitian realizations, the effect of the coordinate noncommutativity is delegated to interactions. Thus, after variation, the deformed Dirac equation generically reads}
\begin{equation}
    (\hatslashed{\pi}-Mc)\psi=0.\label{eqn:DefDirac}
\end{equation}
As we are working in the position representation, the momentum operators stand for $\hk_\mu\psi=i\partial_\mu\psi.$
For the Dirac matrices, we use the Dirac representation
\begin{equation}
\gamma_i=\begin{pmatrix}
0&\sigma_i\\
-\sigma_i&0
\end{pmatrix},\hspace{1cm}\gamma^0=\begin{pmatrix}
\id&0\\
0&-\id
\end{pmatrix},
\end{equation}
with the two-dimensional identity matrix $\id$ and the Pauli matrices $\sigma_i.$

That the MDR given in Eq. \eqref{eqn:mdrkd} is fulfilled implies that the deformed Dirac operator has to obey the condition
\begin{equation}
    {\hpi}^2=-\eta^{\mu\nu}\hpi_\mu\hpi_\nu=\hH^2/c^2-\hk^2+\ell\sum_{n=0}^2\sum_{m=0}^{3-n} a_{n,m}(Mc)^n \hk_\perp^{m}(M^2c^2+\hk_\perp^2)^{\frac{3-n-m}{2}}\equiv\hat{\mathcal{C}}^{(0)}+\ell\hat{\mathcal{C}}^{(1)},\label{eqn:pCond}
\end{equation}
where $\hat{\mathcal{C}}^{(n)}$ are the operator-valued equivalents of the $\mathcal{C}^{(n)}.$
In order for the vector $\hpi_\mu$ to recover $\hk_\mu$ at vanishing scale $\ell,$ it thus has to be of the form
\begin{equation}
    \hpi_\mu=\hk_\mu+\ell\hpi^{(1)}_\mu,
\end{equation}
with the first-order correction $\hpi^{(1)}_\mu$ (of units of $[k^2]$) in $\ell,$ which satisfies the condition
\begin{equation}
    -2\eta^{\mu\nu}\hk_\mu \hpi^{(1)}_\nu=\hat{\mathcal{C}}^{(1)}.\label{eqn:MomCasCorr}
\end{equation}
The most general solution to this equation reads
\begin{equation}
    \hpi^{(1)}_\mu=-\frac{b_{\mu}}{2b^\nu\hk_\nu}\hat{\mathcal{C}}^{(1)},\label{eqn:Defb}
\end{equation}
for some vector $b^\mu.$ 

In order to make the deformed Dirac Hamiltonian manifest, we can single out the time derivative from Eq. \eqref{eqn:DefDirac} and rewrite it as
\begin{equation}
    i\partial_t\psi=c\left(-\gamma^0\gamma_i\hk_i+\gamma^0Mc-\ell\gamma^0\gamma^\mu\hpi^{(1)}_\mu\right)\psi.\label{eqn:AfterPhase}
\end{equation}
In a WKB-like expansion analogous to the one considered in Refs. \cite{Schwartz:2018pnh,Wagner:2023fmb} for scalars and Ref. \cite{Alibabaei:2023dfd} for Dirac spinors, we consider solutions of the form $\psi = e^{-iMc^2t}\tilde{\psi}.$ {There are two possible interpretations of this ansatz: First, it can be understood as a single excitation following an arrow of time pointing towards the future. This time direction will, in the following, become the absolute notion of time required for the nonrelativistic limit. Second, the phase $e^{-iMc^2t},$ for scalar arguments, amounts to a translation in the space of energies, changing the rest mass. As we will see below, its action on the spinor is slightly more involved.

As a result,} the field equation for $\tilde\psi$ becomes
\begin{equation}
    (i\partial_t+Mc^2)\tilde\psi=c\left(-\gamma^0\gamma_i\hk_i+\gamma^0Mc-\ell\gamma^0\gamma^\mu e^{iMc^2t}\hpi^{(1)}_\mu e^{-iMc^2t}\right)\tilde\psi.
\end{equation}
The phase $e^{iMc^2t}$ {acts on functions of the energy by shifting $\hat k_0\to\hat k_0+Mc.$ Thus, its adjoint action on $\hat\pi_\mu^{(1)}$ reads}
\begin{equation}
    e^{iMc^2t}\hat{\pi}_\mu^{(1)}(\hk_0,\hk_i)e^{-iMc^2t}=\hat{\pi}_\mu^{(1)}(\hk_0+Mc,\hk_i)\equiv\hat{\bar{\pi}}_\mu^{(1)}.
\end{equation}
Expressing the spinor field in terms of particle and antiparticle components $\psi=(\varphi,\chi),$ we can rewrite Eq. \eqref{eqn:AfterPhase} as
\begin{equation}
    i\partial_t \begin{pmatrix}
        \varphi\\\chi
    \end{pmatrix}=-c\begin{pmatrix}
        \ell\hbpI_0&\sigma_i(\hk_i+\ell\hbpI_i)\\
        \sigma_i(\hk_i+\ell\hbpI_i)&2Mc+\ell\hbpI_0
    \end{pmatrix}\begin{pmatrix}
        \varphi\\\chi\label{eqn:phichiInput}
    \end{pmatrix}.
\end{equation}
In the nonrelativistic regime, $i\partial_t\chi\ll Mc^2\chi.$ Therefore, we can approximate the field \eqref{eqn:phichiInput} as
\begin{align}
    i\partial_t\varphi=&-c\ell\hbpI_0\varphi-c\sigma_i(\hk_i+\ell\hbpI_i)\chi,\\
    \left(2Mc+\ell\hbpI_0\right)\chi\simeq&-\sigma_i(\hk_i+\ell\hbpI_i)\varphi.\label{eqn:chiFEQ}
\end{align}
{Here, the phase $e^{-iMc^2t}$ has rendered one of the two Weyl spinors subsumed in $\tilde\psi$ (namely $\xi$), say the antiparticle, non-dynamical in the nonrelativistic regime. This is as expected for the nonrelativistic limit, which should be devoid of antiparticles.} At first order in $\ell,$ Eq. \eqref{eqn:chiFEQ} can be approximated as
\begin{equation}
    \chi\simeq-\frac{\sigma_i\left[\left(1-\frac{\ell\hbpI_0}{2Mc}\right)\hk_i+\ell\hbpI_i\right]}{2Mc}\varphi
\end{equation}
The fermion, in turn, satisfies the deformed nonrelativistic Schr\"odinger equation
\begin{align}
    i\partial_t\varphi=&\left[\frac{\left(\sigma_i\hk_i\right)^2+\ell\sigma_i\sigma^j\left(2\hk_{(i}\hbpI_{j)}-\frac{\hbpI_0}{2Mc}\hk_{i}\hk_j\right)}{2M}-\ell\hbpI_0c\right]\varphi,\label{eqn:PlugInChi}\\
    =&\left[\frac{\hk^2\left(1-\ell\frac{\hbpI_0}{2Mc}\right)+2\ell\hk^{i}\hbpI_{i}}{2M}-\ell\hbpI_0c\right]\varphi,\\
\end{align}
where we used the fact that $\hk_i$ and $\hbpI_j$ (products of partial derivatives) commute, and that the product of two Pauli matrices satisfies the identity
\begin{equation}
    \sigma_i\sigma^j=\delta^{ij}\id+i\epsilon^{ijk}\sigma_k.\label{eqn:PauliIdentity}
\end{equation}
Consequently, we obtain the deformed Schr\"odinger equation
\begin{align}
    i\partial_t\varphi=&\left[\frac{\hk^2}{2M}-\frac{\ell\hat{\mathcal{C}}^{(1)}|_{\hk_0\to\hk_0+Mc}}{2M}\frac{-b_0\left(\frac{\hk^2}{4Mc}+Mc\right)+b_i\hk_i}{-b_0(\hk_0+Mc)+b^j\hk_j} \right]\varphi.\label{eqn:ResSchrodp}
\end{align}
Substituting the zeroth-order Schr\"odinger equation in $\ell,$ \ie, $\hk_0\varphi=i\partial_t\varphi/c\simeq \hk^2\varphi/2Mc$, we obtain
\begin{align}
    i\partial_t\varphi=\left(\frac{\hk^2}{2M}-\frac{\ell\hat{\mathcal{C}}^{(1)}|_{\hk_0\to \frac{\hk^2}{2Mc}+Mc}}{2M}\right)\varphi,
   \label{eqn:DefSchrodGen}
\end{align}
unless the highest-order correction in $1/c$ to the Casimir is constant, \ie $\hat{\mathcal{C}}^{(1)}$ is independent of $\hp_\perp.$ As we will not consider these MDRs in the remainder of this article, we treat this case in the appendix.
In the last step leading up to Eq. \eqref{eqn:DefSchrodGen} we have considered only the leading-order term in $1/c$ (\ie, $Mc$), which allowed us to simplify the expression and consequently cancel the dependence on the arbitrary vector $b_\mu$. 

Finally, we {have to specify the nonrelativistic limit} of the correction to the Casimir invariant. Following the reasoning outlined in \cite{Wagner:2023fmb}, we arrive at the expressions 
\begin{align}
    \hk_d|_{\hk_0\to\frac{\hk^2}{2M}+Mc}\varphi\simeq \left(Mc+\frac{\hk_\perp^2}{2Mc}\right)\varphi,&&\hk_\perp^2|_{\hk_0\to\frac{\hk^2}{2M}+Mc}\varphi\simeq \delta^{ij}\left(\hk_i-M\ord{\mathcal{A}}{1}_i\right)\left(\hk_j-M\ord{\mathcal{A}}{1}_j\right)\equiv\hk^2_{\mathcal{A}}\varphi.\label{eqn:DefkA}
\end{align}
For all corrections to the Casimir which depend on $\hk_i$ to highest order (\ie which contain at least one power of $\hk_\perp$), the highest-order approximation in $1/c$ to Eq. \eqref{eqn:ResSchrodp} becomes
\begin{align}
    i\partial_t\varphi=&\frac{\hk^2-\ell\sum_{n=1}^3\xi_n(Mc)^{3-n}\hk^n_{\mathcal{A}}}{2M}\varphi.
\end{align}
Here, we introduced a parametrization already presented in \cite{Wagner:2023fmb}, viz.
\begin{align}
    \xi_1=a_{0,1}+a_{1,1}+a_{2,1},&&\xi_2=a_{0,2}+a_{1,2},&&\xi_3=a_{0,3}.
\end{align}
Thus, for MDRs of the kind presented in Eq. \eqref{eqn:mdr}, whose corrections contain at least one power of $\hk_\perp$, the nonrelativistic dynamics of uncharged particles derived from the Dirac equation (Eq. \eqref{eqn:DefDirac}) and the Klein-Gordon equation (see \cite{Wagner:2023fmb}) coincide.

The particular effects of the spin degrees of freedom are tied to the particle's interaction with the electromagnetic field. We derive the corresponding nonrelativistic dynamics in the following subsection.
 
\subsection{Coupling to the electromagnetic field}

In the following, we demonstrate how the coupling to the electromagnetic field modifies the interaction between the spin and the external magnetic field $B_i$. To clearly identify these corrections, we again take the nonrelativistic limit, focusing on terms involving the coupling $B_i\sigma_i$.

In order to give charge to the spinor, we have to impose $U(1)$ gauge-invariance on the Lagrangian given in Eq. \eqref{eqn:DiracLagNoCharge}. {Typically, this procedure amounts to the minimal-coupling prescription \cite{Bosso:2018uus}
\begin{equation}\label{gc}
    k_\mu\to p^{(0)}_\mu=k_\mu-q A_\mu,
\end{equation}
with the electromagnetic one-form $A_\mu$ and the charge (gauge-coupling) $q$. Throughout this paper we choose to work in the Weyl gauge, \ie $A_\mu=(0,A_i).$

To account for Lorentz-deforming models, we have to generalize the coupling to the electromagnetic field to noncommutative geometries. In comparison to the subtleties of defining consistent gauge theories in Lorentz-deforming scenarios \cite{Mathieu:2020ccc,Hersent:2021sbt,Hersent:2022gry,Vitale:2023znb}\footnote{Indeed, a consistent $U(1)$-gauge theory in $\kappa$-Minkowski spacetime may only be achievable in five dimensions \cite{Mathieu:2020ywc,Hersent:2021esd}.}, this is a rather straightforward exercise.
As the product between position-dependent quantities (such as fields) is deformed on a noncommutative spacetime, a noncommutative analogue to minimal coupling reads
\begin{equation}
    \hat{k}_\mu\psi\to\hat{p}_\mu\psi=\hat{k}_\mu\psi-qA_\mu\SProd\psi,
\end{equation}
where, we recall, $\SProd$ stands for the associative, but not necessarily commutative star product introduced in Eq. \eqref{eqn:KontsevichProd}. Given such a $*$-product, $U(1)$-symmetry amounts to invariance under the local transformation $\psi\to e^{iq\alpha(x)}\SProd\psi,$ $A\to A',$ where $\delta A=A'-A$ is yet to be determined.  

The deformed Dirac-Lagrangian, from which a gauge-invariant version of the field Eq. \eqref{eqn:DefDirac} can be derived, reads
\begin{equation}
    \mathcal{L}=\psi^\dagger\gamma^0\SProd(\hatslashed{\pi}(\hat p)+Mc)\psi,
\end{equation}
where $\hpi$ is an analytic function of $\hat p,$ and $\dagger$ stands for the involution compatible with the star product on the noncommutative geometry. For this Lagrangian to be invariant under $U(1)$-transformations, we have to require that
\begin{equation}
    e^{-iq\alpha(x)}\SProd(\hatslashed{p}-q\delta A_\mu\SProd)e^{iq\alpha(x)}=\hatslashed{p}.\label{eqn:nonpertGaugeCond}
\end{equation}
For infinitesimal transformations, the condition becomes
\begin{equation}
    [\hat{p}_\mu,\alpha]= [\hat{k}_\mu,\alpha]-q[A_\mu,\alpha]_\SProd=\delta A_\mu,
\end{equation}
This implies that the gauge field has to change as
\begin{equation}
    \delta A_\mu=\partial_\mu\alpha-[A_\mu,\alpha]_\SProd
\end{equation}
under a gauge transformation to satisfy local $U(1)$-symmetry.}

Generally, the coupling to the electromagnetic field induces noncommutativity in the momenta, namely to order $\ell$
\begin{equation}
    \left[\hp_\mu,\hp_\nu\right]\psi\simeq i(qF_{\mu\nu}\SProd+2q^2\bar\Theta^{\rho\sigma}\partial_\rho A_\mu\partial_\sigma A_\nu)\psi\label{eqn:DefEMp}
\end{equation}
with the electromagnetic field-strength tensor for vanishing noncommutativity $F_{\mu\nu}=2\partial_{[\mu}A_{\nu]}.$ {As we only apply the commutator of physical momenta at order $\ell$, we approximate it as
\begin{equation}
    \left[\hp_\mu,\hp_\nu\right]\psi\simeq iqF_{\mu\nu}\psi.\label{eqn:DefEM}
\end{equation}
}In the following, we will expand our results to linear order in the gauge coupling, assuming it to be small,\footnote{From the phenomenological perspective, it is even harder to detect Planck-suppressed terms, which are further undermined by high powers of the gauge coupling (recall the fine-structure constant has the value $\alpha_{U(1)}\simeq1/137$).} and neglect terms containing more than one commutator. 

In contrast to the derivation carried out above, we need to keep track of the ordering at every step of the derivation. Starting with the Casimir invariant in the absence of coordinate noncommutativity (which we assume to be symmetric), we obtain
\begin{equation}
    \hat{\mathcal{C}}^{(1)}|_{\bar\Theta=0}=\sum_{n=0}^2\sum_{m=0}^{3-n}a_{n,m}(Mc)^n\mathcal{S}\left[(\hp^{(0)}_\perp)^m (\hp^{(0)}_d)^{3-m-n}\right],
\end{equation}
where $\mathcal{S}$ stands for a general symmetric ordering prescription underlying the model at hand. Then, the correction to the gauge-covariant Dirac operator $\hPi_\mu^{(1)}$ does not satisfy Eq. \eqref{eqn:MomCasCorr}, but rather
\begin{equation}
    -\eta^{\mu\nu}\{\hPi_\mu^{(1)},\hp^{(0)}_\nu\}=-\eta^{\mu\nu}\left(2\hPi_\mu^{(1)}\hp^{(0)}_\nu+[\hp^{(0)}_\mu,\hPi_\nu^{(1)}]\right)=\hat{\mathcal{C}}^{(1)}|_{\Theta^{\mu\nu}=0},\label{eqn:DiracConCov}
\end{equation}
{where it is understood that commutators are evaluated at vanishing noncommutativity because contributions dependent on $\bar\Theta^{\mu\nu}$ are of higher order in $\ell.$} As $\hp^{(0)}_\mu$ and $\hat{\mathcal{C}}^{(1)}|_{\Theta^{\mu\nu}=0}$ are both symmetric, the Hermitian conjugate of the Dirac operator also satisfies Eq. \eqref{eqn:DiracConCov}. In other words, the gauge-covariant Dirac operator is symmetric. Let us expand the correction to the Dirac operator and the quantity $\hat{\mathcal{C}}^{(1)}|_{\Theta^{\mu\nu}=0}$ in $q$, such that 
\begin{equation}
    \hPi_\mu^{(1)}=\sum_{n=0}^\infty q^{n}\hPi_\mu^{(1,n)}\,, \qquad \hat{\mathcal{C}}^{(1)}|_{\Theta^{\mu\nu}=0}=\sum_{n=0}^\infty q^n \hat{\mathcal{C}}^{(1,n)}|_{\Theta^{\mu\nu}=0}\,.
\end{equation}
Then, to leading order in $q,$ we can neglect the commutator in Eq. \eqref{eqn:DiracConCov} to obtain the solution
\begin{align}
    \hPi_\mu^{(1,0)}=-\frac{1}{2}\hat{C}^{(1,0)}|_{\Theta^{\mu\nu}=0}(b^\nu\hk_\nu)^{-1}b_\mu.
\end{align}
As a first iterative step, Eq. \eqref{eqn:DiracConCov} at order $q$ becomes
\begin{equation}
    -\eta^{\mu\nu}\left(2\hPi_\mu^{(1,1)}\hk_\nu+[\hk_\mu,\hPi^{(1,1)}_\nu]\right)=\hat{\mathcal{C}}^{(1,1)}|_{\Theta^{\mu\nu}=0}-\eta^{\mu\nu}\left(2\hPi_\mu^{(1,0)}A_\nu+[A_\mu,\hPi^{(1,0)}_\nu]\right).
\end{equation}
The solution to this equation would provide the analytic expression of the term $\hPi_\mu^{(1,1)}$.
However, as in general all commutators in the above expression are non-vanishing and since $\hat{\mathcal{C}}^{(1,1)}|_{\Theta^{\mu\nu}=0}$ is in principle a function of both $\hk_\mu$ and $A_\mu$, a general solution cannot be found in a straightforward way. Nevertheless, in the examples we consider in the subsequent sections, the shape of $\hPi_\mu^{(1,1)}$ can be determined straightforwardly.

Thus, the deformed Schr\"odinger equation for nonrelativistic charged spinors reads
\begin{align}
    i\partial_t\varphi =& \left.\left[\frac{\left(\sigma_i\hp_i\right)^2+\ell\sigma_i\sigma^j\left(\hp_{i}\hPi^{(1)}_{j}+\hPi^{(1)}_{i}\hp_{j}-\frac{1}{2Mc}\hp_{i}\hPi_0^{(1)}\hp_j\right)}{2M}-\ell\hPi_0^{(1)}c\right]\varphi\right|_{\hk_0\varphi\to Mc\varphi},\\
    =& \left.\left[\frac{\hp^2-q\sigma_iB_i\SProd}{2M}+\frac{\ell}{2M}\left(\{\hp_i,\hPi^{(1)}_i\}-\frac{\hp^{i}\hPi^{(1)}_0\hp_i}{2Mc}+i\epsilon^{ijk}\sigma_i\left([\hp_j,\hPi^{(1)}_k]-\frac{\hp_j\hPi^{(1)}_0\hp_k}{2Mc}\right)\right)-\ell\hPi_0c\right]\varphi\right|_{\hk_0\varphi\to Mc\varphi},
    \label{eqn:ToEvaluateInCases}
\end{align}
where we used Eq. \eqref{eqn:PauliIdentity}, and defined the magnetic field as $B_i\equiv \epsilon^{ijk}\partial_jA_k=\epsilon^{ijk}F_{jk}/2$. 

{The contributions of the coordinate noncommutativity to Eq. \eqref{eqn:ToEvaluateInCases} are hidden in the first term on the right-hand side. In particular, we can write
\begin{align}
    \left(\hp^2-q\sigma_iB_i\SProd\right)\varphi\simeq&\left[(\hp^{(0)})^2-q\sigma_iB_i-\frac{q\bar\Theta^{\mu\nu}}{2}\left(2\partial_{\mu}A_i\hk_i\hk_{\nu}+\sigma_i\partial_\mu B_i\hk_\nu\right)\right]\varphi+\mathcal{O}(q^2),    \label{eqn:NoncommContSchrod}
\end{align}
where $\hp^{(0)}_i=\hp_i|_{\bar\Theta=0},$ and we considered that magnetic fields have no sources to lowest order in $\ell$, \ie $\partial_iB_i=0.$
In a specific model with given $\bar\Theta^{\mu\nu}$, we can evaluate these corrections explicitly and compare to corrections included by the deformation of the Dirac operator to obtain the most relevant contributions in the nonrelativistic limit.

In this paper we aim to specify corrections to the spin coupling. Therefore, in the following Section, where we provide some notable example models, we are only interested in those terms which are proportional to the spin operator $\sigma_i/2$ in Eq. \eqref{eqn:ToEvaluateInCases}.}

\section{Model examples}
\label{sec:examples}

In this Section, we explicitly analyze several relevant example models to investigate the deformation to the Pauli equation and point out the ensuing corrections to the spin-coupling terms. In particular, we consider two representations of the $\kappa$-Poincar\'e algebra, the classical basis and the bi-crossproduct basis, and two common models of minimal-length quantum mechanics.

{
\subsection{Classical representation of the $\kappa$-Poincar\'e algebra\label{sec:ClassRep}}

The $\kappa$-Poincar\'e group \cite{Majid:1994cy,Sitarz:1994rh} is one of the most commonly studied models of deformed spacetime symmetry transformations in quantum gravity phenomenology \cite{Kowalski-Glikman:2002iba,Kowalski-Glikman:2002eyl,Bruno:2001mw} (see \cite{Arzano:2021scz} for a recent monograph).  It deforms the Poincar\'e group to accommodate Lie-type coordinate noncommutativity, with the deformation governed by the scale $1/\kappa \propto \ell$, governing non-linearities in momentum space \cite{Kowalski-Glikman:2002oyi,Kowalski-Glikman:2003qjp}. Distinct representations of the $\kappa$-Poincar\'e algebra exist \cite{Kowalski-Glikman:2002iba}, related by active diffeomorphisms in momentum space \cite{Amelino-Camelia:2019dfl}. While the (original) bi-crossproduct basis \cite{Majid:1994cy} is the focus of the next Subsection, this Subsection examines the classical basis, which is widely used in noncommutative field theories \cite{Arzano:2009ci,Kowalski-Glikman:2009len,Meljanac:2010ps,Meljanac:2011cs,Arzano:2017uuh,Arzano:2018gii,Arzano:2020jro,Bevilacqua:2022fbz,Bevilacqua:2024jpy,Bevilacqua:2024qdh}.

The classical basis is defined such that the Dirac operator is undeformed. As a result, modifications to the Pauli equation can only surface through the interaction with the magnetic field. The hermitian star product reads to first order in $\ell$ \cite{Meljanac:2010ps}
\begin{equation}
    f\star g\simeq fg+i\ell x^\mu(-\partial_\mu f \partial_dg+d_\mu\partial_\lambda f\partial^\lambda g).
\end{equation}
This implies that
\begin{equation}
    \bar\Theta^{\mu\nu}=2\ell(-x^\mu d^\nu+x^\rho d_\rho\eta^{\mu\nu}).
\end{equation}
As a result, the effective the Pauli equation reads
\begin{align}
    i\partial_t\varphi=\frac{1}{2M}&\left.\left[(\hat p^{(0)})^2-q\sigma_iB_i-q\ell\left(-x^\mu d^\nu+x^\rho d_\rho\eta^{\mu\nu}\right)(2\partial_\mu A_i\hat k_i\hat k_\nu+i\partial_\mu\partial_i A_i\hat k_\nu+\sigma_i\partial_\mu B_i\hat k_\nu)\right]\varphi\right|_{\hat k_0\varphi\to Mc\varphi}\\
    \simeq\frac{1}{2M}&\left.\left[(\hat p^{(0)})^2-q\sigma_iB_i-iq\ell Mc\left(2[\hat D,A_i]\hk_i+[\hat D,\partial_iA_i]+\sigma_i [\hat D,B_i]\right)\right.\right.\nonumber\\
    &\left.\left.-q\ell tc\left(\partial^\mu (2A_i\hk_i+\partial_iA_i+\sigma_iB_i)\hk_\mu\right)\right]\varphi\right|_{\hk_0\varphi\to Mc\varphi},
\end{align}
where we introduced the dilatation generator $\hat D\varphi=ix^\mu\partial_\mu\varphi.$ Expressed in this way, we find that many of the corrections amount to a scale transformations of the involved fields, \eg for a generic field $\Phi(x,t)$
\begin{equation}
    \Phi+i\ell Mc[\hat D,\Phi]\simeq e^{i\ell Mc\hat D}\Phi(x,t)e^{-i\ell Mc}=\Phi\left[\left(1+\ell M c\right)x,\left(1+\ell M c\right)t\right]\equiv \Phi^{\hat{D}}(x,t),
\end{equation}
where the last equality defines the dilated field $\Phi^{\hat D}.$ Consequently, the corrected Pauli equation reads to leading order in $\ell$ and $c^{-1}$
\begin{equation}
    i\partial_t\varphi=\frac{(\hp^{(0)})^2|_{\Phi=\Phi^{\hat D}}-q\sigma_i B^{\hat D}_i}{2M}\varphi-q\ell tc\partial^\mu (2A_i\hk_i+\partial_iA_i+\sigma_iB_i)\hk_\mu\varphi|_{\hk_0\varphi\to Mc\varphi},\label{eqn:k-PoincareSchrodClass}
\end{equation}
where $\Phi$ stands for any function of positions involved, \ie $\Phi=(A_i,\partial_jA_j).$ 

When the background vector potential or magnetic field is position-dependent, the Hamiltonian becomes explicitly time-dependent. As a result, the system is effectively open, with Planck-scale corrections breaking energy conservation. Note that the time dependence of the Hamiltonian is parametric because the background dynamics are not derived from a quantum-mechanical model. A more realistic, fully quantum treatment would induce fundamental decoherence \cite{Arzano:2022nlo} (see also \cite{Petruzziello:2020wkd,DEsposito:2023psn}), offering an additional avenue for high-precision experimental tests \cite{Domi:2024ypm}.

The coupling between spin and magnetic field, which we concentrate most on in this paper, reads
\begin{equation}
    \hH_{\rm spin}=-\frac{q\sigma_i}{2M} \left[B^{\hat D}_i+q\ell t c(\partial_jB_i\hk_j-M\partial_t B_i)\right].
\end{equation}
Note that Planck-scale corrections to this coupling vanish in homogeneous and static magnetic fields to all orders in $\ell$, as they arise exclusively from the star product. We now turn to the bi-crossproduct representation.
}

\subsection{Bi-crossproduct representation of the $\kappa$-Poincar\'e algebra\label{sec:Bicross}}

{The bi-crossproduct basis was the original formulation of the $\kappa$-Poincar\'e algebra \cite{Majid:1994cy} and has since been widely studied in the context of noncommutative field theory \cite{Amelino-Camelia:2001rtw,Poulain:2018mcm,Poulain:2018two,Mathieu:2020ywc,Mathieu:2020ccc,Mathieu:2021mxl,Hersent:2021esd,Hersent:2021sbt,Hersent:2022gry,Hersent:2020lsr}, where it has been shown that scalar field theories are indeed invariant under deformed versions of translations and Lorentz transformations \cite{Poulain:2018mcm}. Here we assume that this property carries over to Dirac fields. 

Several derivations of the deformed Dirac operator in the bi-crossproduct representation exist \cite{Nowicki:1992if,Bibikov:1997qu,Sorace:2000kal,DAndrea:2005hjg,Poulain:2018mcm}. Here, we employ an approach designed to ensure that the Klein-Gordon operator equals the square of the Dirac operator, which transforms covariantly under the deformed Lorentz transformations \cite{Franchino-Vinas:2022fkh}.}

Recently, deformations of spacetime symmetries have been fully translated into a momentum-space geometric picture \cite{Gubitosi:2011hgc,Carmona:2019fwf,Relancio:2020rys,Relancio:2020zok,Relancio:2021ahm,Carmona:2021gbg,Franchino-Vinas:2023rcc,Relancio:2024axb}, which has been applied to the Dirac equation in \cite{Franchino-Vinas:2022fkh}. Accordingly, we can associate to a quantum spacetime a momentum dependent metric $\hat{g}^{\mu\nu}(\hk)$ and a momentum dependent tetrad $\hat{e}_a^{~\mu}$ such that the length element
\begin{equation}
    \D s^2=\hat{g}^{\mu\nu}(\hk)\D \hk_\mu\D \hk_\nu=\hat{e}_{~a}^{\mu}(\hk)\eta^{ab}\hat{e}_b^{~\nu}(\hk)\D \hk_\mu\D \hk_\nu
\end{equation}
assigns distances in momentum space. The Casimir of the $\kappa$-Poincar\'e algebra $\mathcal{C}$ can then be understood as the geodesic distance from the origin in momentum space \cite{Gubitosi:2011hgc}, thus satisfying \cite{Carmona:2019fwf}
\begin{equation}
    \hat{\mathcal{C}}=-\eta^{ab}\hpi_a\hpi_b,\label{eqn:MomGeoDist}
\end{equation}
with $\hpi_a=-e_{a\mu}\mompar^\mu\hat{\mathcal{C}}/2,$ and where $\mompar^\mu=\partial/\partial k_\mu.$ 

The Casimir invariant yields the MDR exactly as given in Eq. \eqref{eqn:mdr}. Bearing this fact in mind, Eqs. \eqref{eqn:pCond} and \eqref{eqn:MomGeoDist} are indeed equivalent, with the operators $\hpi_a$ and $\hpi_\mu$ playing the same role. Then, Eq. \eqref{eqn:DefDirac} applies equivalently to the present case if $\hatslashed{\pi}=\gamma^a\hpi_a,$ where again the properties of the matrices $\gamma^a$ are the same as the properties of the matrices $\gamma^\mu.$

In the bi-crossproduct representation, the Casimir invariant and the tetrad read \cite{Carmona:2019fwf}
\begin{align}
    \hat{\mathcal{C}}=&\frac{2}{(\lambda\ell)^2}\left(1-\cosh{\lambda\ell\hk_d}\right)-e^{\lambda\ell\hk_d} \hk^2\simeq \hk_d^2-\left(1+\lambda\ell\hk_d\right)\hk_\perp^2, && \hat{e}^a_{~\mu}=\begin{pmatrix}
        1 & 0\\
        0 & e^{\lambda\ell\hk_d}\delta_{ij}
        \end{pmatrix}\simeq \delta^a_\mu+\lambda\ell\hk_d\begin{pmatrix}
        0 & 0\\
        0 & \delta_{ij}
        \end{pmatrix},
\end{align}
with the dimensionless model parameter $\lambda$ such that the energy scale $\kappa=c(\lambda\ell)^{-1}$ (which gives the model its name) characterizes its effects. Note that $\kappa$-Poincar\'e models by construction do not have a preferred frame such that the estimate on $\ord{\mathcal{A}}{1}_i$ given in Eq. \eqref{eqn:EstimatedDeformation} does not apply. Instead, it is commonly assumed that $\ord{A}{1}_i=0.$

Consequently, the deformation of the momentum operator becomes at first order
\begin{equation}
    \hpi_a^{(1)}=\frac{\lambda}{2}\hk_\perp^2\delta_a^0.
\end{equation}
In other words, it is given by Eq. \eqref{eqn:MomCasCorr} with $b_\mu\D x^\mu=\D t.$ As a result, the deformed Schr\"odinger equation for uncharged fermions in the bi-crossproduct basis of the $\kappa$-Poincar\'e algebra reads
\begin{align}
    i\partial_t\varphi=&\frac{\hk^2-\lambda \ell Mc\hk_{\mathcal{A}}^2}{2M}\varphi\xrightarrow{\ord{\mathcal{A}}{1}_i\to0}\frac{\hk^2}{2M}\left(1-\lambda\ell Mc\right)\varphi=\frac{\hk^2}{2\bar{M}}\varphi,
\end{align}
which effectively amounts to a change in the inertial mass
\begin{equation}
    \bar{M}=M(1+\lambda \ell Mc)
\end{equation}
exactly as predicted for scalars in \cite{Wagner:2023fmb}. 

{To couple the particle to the electromagnetic field, we need to supplement the Casimir with a suitable $*$-product. In the bi-crossproduct representation, the star product reads to first order in $\ell$ \cite{Majid:1994cy,Kowalski-Glikman:2002iba,Kowalski-Glikman:2002eyl,Meljanac:2007xb} 
\begin{equation}
    f\star g=fg-i\ell x^\mu(\delta_\mu^\nu-d_\mu d^\nu)d^\rho\partial_\rho f\partial_\nu g\simeq fg+\mathcal{O}(\ell c^{-1}).
\end{equation}
The contribution from the noncommutative geometry is subleading in $c^{-1}$ compared to the effective modification of the inertial mass in the kinetic term. Notably, some of these subleading contributions, as in the classical basis, would be explicitly time-dependent. This again suggests a decohering influence of the background when its quantum nature is fully taken into account.

Similarly, all additional corrections stemming from the coupling of the particle to the magnetic field are negligible in the nonrelativistic regime. Consequently, a fermion carrying a $U(1)$-charge satisfies the deformed Pauli equation
\begin{align}
    i\partial_t\varphi=&\left(\frac{\hp^2-\lambda \ell Mc\hp_{\mathcal{A}}^2}{2\bar{M}}-\frac{qB_i\sigma_i}{2M}\right)\varphi\xrightarrow{\ord{\mathcal{A}}{1}_i\to0}\left(\frac{\hp^2}{2\bar{M}}-\frac{qB_i\sigma_i}{2M}\right)\varphi.\label{eqn:k-PoincareSchrod}
\end{align}
While the inertial mass is effectively deformed, the coupling to the magnetic field retains its dependence on $M$. The Hamiltonian describing the interaction of magnetic field and spin reads
\begin{equation}
    \hat H_{\rm spin}\psi=-\frac{\sigma_iB_i}{2M}\psi=-\frac{\sigma_iB_i}{2\bar M}(1+\lambda\ell\bar M c)\psi,
\end{equation}
as in ordinary Galilean quantum theory. Corrections are hidden in the differing masses in kinetic term and spin interaction.}

\subsection{Minimal-length quantum mechanics\label{sec:MinLen}}

{
The most commonly used model of minimal-length quantum mechanics, also referred to as the commutative Euclidean Snyder model, assumes vanishing coordinate noncommutativity and yields $\hat{\mathcal{C}} = \mathcal{C}^{(0)} - 2\beta\ell^2\hk_{\perp}^4$, where $\beta$ is the dimensionless model parameter. Notably, unlike the previous example, the leading-order contribution in this model is proportional to $\ell^2$. However, the first-order corrections of interest are computed in the same way, regardless of the power of $\ell$ suppressing them.}

Consequently, the leading-order deformation of the momentum $\hpi^{(2)}_\mu$ (now having units of $[k^3]$) is given as
\begin{equation}
    \hpi^{(2)}_\mu=\beta\hk^2\left(\hk_\mu-\hk_dd_\mu\right)\,.
\end{equation} 
As indicated previously, when considering the gauge coupling in Eq. \eqref{gc} and the nonrelativistic limit, only the terms which are contracted with the Levi-Civita symbol are relevant for our purposes, since it is possible to extract the coupling between the spin operator $\hat{S}_i=\sigma_i/2$ and the magnetic background only from such factors.
Hence, in the following we are only interested in the part of the nonrelativistic wave equation that reads
\begin{equation}
    i \epsilon_{ijk} \left(\hp_i\hp_j+\beta\ell^2\hp_i\hp_\perp^2\hp_j+\beta\ell^2\hp_\perp^2\hp_i\hp_j\right)\sigma_k.
\end{equation}
Under the assumption that $\ord{\mathcal{A}}{1}_i$ is effectively homogeneous within the experimental setup\footnote{This implies that the background deformation vector changes only on scales much larger than the lab, which an experiment is done in.} and applying Eq. \eqref{eqn:DefEM}, we rewrite these modifications as 
\begin{equation}
    -(1+2\beta\ell^2\hp_\perp^2)qB_i\sigma_i+q\beta\ell^2\epsilon_{ijk}\left\{F_{\mu i},\hp^\mathcal{A}_l\right\}\hp_j\sigma_k,
\end{equation}
where $\hp^{\mathcal{A}}_i\equiv\hp_i-M\ord{\mathcal{A}}{1}_i.$ By manipulating the second contribution in the above expression, we further obtain
\begin{equation}
q\beta\ell^2\epsilon_{ijk}\left(2\hp^{\mathcal{A}}_l\hp_jF_{li}-\left[\hp_l\hp_j,F_{li}\right]-\hp_l\left[\hp_j,F_{li}\right]\right)\sigma_k=-2q\beta\ell^2\left(\hp^{\mathcal{A}}_j\hp_jB_i\sigma_i-\hp_i\sigma_i\hp^{\mathcal{A}}_jB_j\right),\label{eq:GUPCorrSimp}
\end{equation}
where we assumed the magnetic field $B_i$ to be homogeneous and static in the last step.
Although this assumption has been introduced for the sake of streamlining calculations, it does not affect our phenomenological conclusions in Section \ref{sec:anomalous_magmom}.

By accounting for all the above caveats and focusing the attention only on the terms deemed relevant for the current analysis, it is possible to write the deformed Schr\"odinger equation for the present model as
\begin{equation}
    i\partial_t\varphi = \left[\left(1+2\beta\ell^2\hp_{\mathcal{A}}^2\right)\frac{\hp^2}{2M} -\left(1+2\beta\ell^2(\hp_{\mathcal{A}}^2+\hp^{\mathcal{A}}_i\hp_i)\right)\frac{qB_i\sigma_i}{2M}+\frac{\beta\ell^2q}{2M}\hp_i\sigma_i\hp^{\mathcal{A}}_jB_j\right]\varphi.\label{eqn:quadGUPSchrod}
\end{equation}
In common models $\ord{\mathcal{A}}{1}_i$ is assumed to vanish such that the corrections become
\begin{equation}
    i\partial_t\varphi = \left[\left(1+2\beta\ell^2\hp^2\right)\frac{\hp^2}{2M} -\left(1+4\beta\ell^2\hp^2\right)\frac{qB_i\sigma_i}{2M}+\frac{\beta\ell^2q}{2M}\hp_i\sigma_i\hp_jB_j\right]\varphi.\label{eqn:quadGUPSchrodA0}
\end{equation}
Note that, while the corrections to the spin-magnetic coupling appear non to to be Hermitian, {they are so for homogeneous magnetic fields, which we have assumed in Eq. \eqref{eq:GUPCorrSimp}.} Consequently, the coupling of spin and magnetic field reads
\begin{equation}
    \hat H_{\rm spin}=-\left(1+2\beta\ell^2(\hp_{\mathcal{A}}^2+\hp^{\mathcal{A}}_i\hp_i)\right)\frac{qB_i\sigma_i}{2M}+\frac{\beta\ell^2q}{2M}\hp_i\sigma_i\hp^{\mathcal{A}}_jB_j\xrightarrow{\ord{\mathcal{A}}{1}_i\to0}-\left(1+4\beta\ell^2\hp^2\right)\frac{qB_i\sigma_i}{2M}+\frac{\beta\ell^2q}{2M}\hp_i\sigma_i\hp_jB_j.\label{eqn:quadGUPIntHam}
\end{equation}

Analogous considerations apply to the linear version of minimal-length quantum mechanics, for which $\hat{\mathcal{C}}^{(1)}=2\alpha\hk_\perp^3$, $\alpha$ being the dimensionless parameter of the model. In this case, the deformed momentum operator reads
\begin{equation}
    \hpi^{(1)}_\mu=-\alpha|\hk_\perp|\left(\hk_\mu-\hk_d\delta^d_\mu\right)\,.
\end{equation}
The derivation of the nonrelativistic limit proceeds in a straightforward way and in close analogy to the previous example. Hence, we infer that the nonrelativistic wave equation reads
\begin{equation}
    i\partial_t\varphi=\left[\left(1-2\alpha\ell|\hp_{\mathcal{A}}|\right)\frac{\hp^2}{2M}-\left(1-2\alpha\ell\left(|\hp_{\mathcal{A}}|+\frac{\hp^{\mathcal{A}}_j\hp_j}{|\hp_{\mathcal{A}}|}\right)\right)\frac{qB_i\sigma_i}{2M}-\frac{\alpha\ell q}{2M|\hp_{\mathcal{A}}|}\hp_i\sigma_i\hp^{\mathcal{A}}_jB_j\right]\varphi,\label{eqn:linGUPSchrod}
\end{equation}
which in the limit $\ord{\mathcal{A}}{1}_i=0$ becomes
\begin{equation}
    i\partial_t\varphi=\left[\left(1-2\alpha\ell|\hp|\right)\frac{\hp^2}{2M}-\left(1-4\alpha\ell|\hp|\right)\frac{qB_i\sigma_i}{2M}-\frac{\alpha\ell q}{2M\hp}\hp_i\sigma_i\hp_jB_j\right]\varphi.\label{eqn:linGUPSchrodA0}
\end{equation}
The corresponding interaction Hamiltonian of spin and magnetic field reads
\begin{equation}
    \hat H_{\rm spin}=-\left(1-2\alpha\ell\left(|\hp_{\mathcal{A}}|+\frac{\hp^{\mathcal{A}}_j\hp_j}{|\hp_{\mathcal{A}}|}\right)\right)\frac{qB_i\sigma_i}{2M}-\frac{\alpha\ell q}{2M|\hp_{\mathcal{A}}|}\hp_i\sigma_i\hp^{\mathcal{A}}_jB_j\xrightarrow{\ord{\mathcal{A}}{1}_i\to0}-\left(1-4\alpha\ell|\hp|\right)\frac{qB_i\sigma_i}{2M}-\frac{\alpha\ell q}{2M\hp}\hp_i\sigma_i\hp_jB_j,\label{eqn:linGUPIntHam}
\end{equation}
thus concluding our considerations on effective nonrelativistic minimal-length models.

With the modified Pauli equations given in Eqs. \eqref{eqn:k-PoincareSchrodClass}, \eqref{eqn:k-PoincareSchrod}, \eqref{eqn:quadGUPSchrod}, and \eqref{eqn:linGUPSchrod} for the classical and bi-crossproduct representations of the $\kappa$-Poincar\'e model, as well as for quadratic and linear minimal-length quantum mechanics, we can directly infer novel phenomenology. For instance, in the next section, we predict modifications to the electron’s anomalous magnetic moment.

\section{The anomalous magnetic moment of the electron}
\label{sec:anomalous_magmom}

In this section, we compare the predictions we derived from Planck-scale modifications to relativistic kinematics with Penning-trap measurements of the electron’s anomalous magnetic moment. Since these experiments are conducted in homogeneous magnetic fields, no constraints are imposed on the classical basis of $\kappa$-Poincar\'e geometry introduced in Section \ref{sec:ClassRep}.

The spin magnetic moment is defined such that the Hamiltonian of the spin coupling reads
\begin{equation}
    H_{\rm spin}=\mu_{S,i}B_i.
\end{equation}
In homogeneous magnetic fields, $\mu_{S,i}$ is deformed as
\begin{equation}
    \mu_{S,i} = - \left[1+f\left(p,M\right)\right] g_S\mu_BB_i,
\end{equation}
with the Bohr magneton $\mu_B=q/2M$, the Land\'e factor $g_S$, and the function $f(p,M)$ which depends on the given deformation of MDRs. This can be understood as a modification of $g_s$ itself, inasmuch as
\begin{equation}
    g_S\to\left(1+f\left(p,M\right)\right)g_S.
\end{equation}
Let us now estimate the strength of this effect by resorting to Penning trap experiments. For semiclassical states describing particles at velocities $v\ll |\ord{\mathcal{A}}{1}_i|\sim10^{-3}c$ (recall Eq. \eqref{eqn:EstimatedDeformation}), as achieved in Penning traps \cite{Brown:1985rh,Dehmelt:1988ga,Hanneke:2008tm,Hanneke:2010au}, we can approximate
\begin{equation}
    \braket{p^n}\simeq(M |\ord{\mathcal{A}}{1}_i|)^n.
\end{equation}
If, instead, $\ord{\mathcal{A}}{1}_i=0$ by assumption, we estimate
\begin{equation}
    \braket{p^n}\simeq(Mv)^n.
\end{equation}

In recent Penning trap measurements \cite{Hanneke:2008tm,Hanneke:2010au}, electrons were slowed down to a velocity of $v=10^{-5}c.$ Furthermore, the mass of the electron relatively to the Planck scale is of the order $\ell Mc\sim 10^{-22}.$ 

{
The magnetic moment of the electron has been predicted and measured with an accuracy of 13 digits, allowing the free parameters in the minimal-length models introduced in Section \ref{sec:MinLen} to be tightly constrained, outperforming many existing bounds in the literature (for a summary, see \cite{Bosso:2023aht}). For nonvanishing $\ord{\mathcal{A}}{1}_i,$ we find
\begin{align}\label{boundsMinLenAneq0}
\sqrt{\beta}\lesssim10^{15},&&\alpha\lesssim 10^{12},
\end{align}
corresponding to the energy scales
\begin{align}
    c(\sqrt{\beta}\ell)^{-1}\gtrsim 10^4\rm{GeV},&&c(\alpha\ell)^{-1}\gtrsim 10^{7}\rm{GeV}.
\end{align}
If we instead assume that $\ord{\mathcal{A}}{1}_i=0,$ the constraints weaken slightly to
\begin{align}\label{boundsMinLenAneq0}
\sqrt{\beta}\lesssim10^{19},&&\alpha\lesssim 10^{16}.
\end{align}
with the associated energy scales
\begin{align}
    c(\sqrt{\beta}\ell)^{-1}\gtrsim 1\rm{GeV},&&c(\alpha\ell)^{-1}\gtrsim 10^{5}\rm{GeV}.
\end{align}
These results emphasize that conventional minimal-length models, involving parameters $\alpha$ and $\beta$, are more effectively tested in relativistic systems, even though the existing literature has predominantly focused on the Galilean regime.

For the bi-crossproduct representation of the $\kappa$-Poincar\'e algebra introduced in Section \ref{sec:Bicross}, our result is independent of the value of the deformation vector. As a result, we obtain
\begin{align}
    \lambda\leq10^9,&&\kappa=c(\lambda\ell)^{-1}\geq 10^{10}\rm{GeV}.
\end{align}
As in \cite{Wagner:2023fmb}, we find that the modifications modeled by the $\kappa$-Poincar\'e deformation are more suited to nonrelativistic phenomenology than minimal-length models.}
Concerning the bound on $\lambda$, we note that the free parameter appearing in different models of $\kappa$-Poincar\'e involving spatial noncommutativity has been constrained with a high degree of accuracy with astrophysical observations \cite{Ellis:2005sjy,FermiGBMLAT:2009nfe,HESS:2011aa,Vasileiou:2013vra,MAGIC:2020egb,LHAASO:2024lub}, albeit with possibly large systematic uncertainties. Our results thus highlight the complementarity of astrophysical observations and table-top experiments.


\section{Discussion}
\label{sec:conclusions}

{
In quantum gravity phenomenology, models with MDRs and noncommutative geometries offer one of the few testable effects at accessible energy scales with near-Planckian sensitivity. These models arise from deformed symmetries that introduce an additional invariant scale into special relativity. Prominent examples include the $\kappa$-Poincar\'e symmetry group, leading to $\kappa$-Minkowski spacetime, and Snyder models, which incorporate a minimal length into quantum mechanics.

While the high energies in astrophysical observations present an obvious amplification of quantum-gravitational effects, more subtle amplifiers can be found in the table-top regime, allowing for precise control over the studied system. These amplifiers may, \eg, be measurement precision or coherence time, thus yielding competitive constraints \cite{Amelino-Camelia:2009wvc,Wagner:2023fmb,Hohmann:2024lys}. Low-energy experiments are best conceived in the language of single-particle quantum mechanics.

In this study, we have focused on the implications of deformed relativistic kinematics on charged fermions. We have derived the nonrelativistic limit of a general deformed Dirac equation incorporating both a general MDR as well as a noncommutative geometry. We have displayed the resulting deformed Pauli equation, and in particular the coupling between spin and magnetic field, for several widely employed models: the $\kappa$-Poincar\'e algebra in both the classical basis and the bi-crossproduct basis, as well as the two pertinent models of minimal-length quantum mechanics (also known as commutative Euclidean Snyder models).

We have found that, except for the classical representation of the $\kappa$-Poincar\'e algebra, the coupling of particles to the magnetic field is generically deformed by terms dependent on the characteristic scales of the model. Using the anomalous magnetic moment of the electron, measured with remarkable accuracy, we have derived bounds on these scales. For example, in the bi-crossproduct basis of the $\kappa$-Poincar\'e algebra, we have obtained $\kappa \geq 10^{10}\mathrm{GeV}$, far exceeding energy scales accessible in current colliders. Although these results do not achieve Planckian precision as astrophysical observations do, they underscore the relevance of the nonrelativistic regime in quantum gravity phenomenology, encouraging further exploration and refinement.
}

In particular, the anomalous magnetic moment of the electron is to be understood solely as a first application, where measurement precision plays the role of an amplifier. More intricate proposals to test tentative quantum gravity-induced modifications (expanding the use of amplifiers) will be the subject of future research.

\appendix

\section{Deformed Schr\"odinger equation for MDRs independent of $\hk_\perp$}

As indicated in Sec. \ref{sec:spin_coupling}, for MDRs whose corrections are independent of $\hp_\perp$, $\mathcal{C}^{(1)}(Mc,\hk_i)$ is constant, and thus the relativistic correction in Eq. \eqref{eqn:DefSchrodGen} is trivial. This implies that physically relevant corrections appear only to higher order in $1/c$. As a result, we obtain the deformed Schr\"odinger equation
\begin{align}
    i\partial_t\varphi=\frac{\hk^2-\bar{\xi}_2\ell M'c \hk_{\mathcal{A}}^2}{2M'}\varphi,
\end{align}
where the inertial mass is modified as
\begin{equation}
    M'=\begin{cases}
        M&\text{if }b_0= 0,\\
        M\left(1-\frac{\ell Mc}{4}\sum_{n=0}^{2}a_{n,0}\right)&\text{if }b_0\neq 0,
    \end{cases}
\end{equation}
and we defined the parameter
\begin{equation}
    \bar{\xi}_2=\frac{3}{2}a_{0,0}+a_{1,0}+\frac{1}{2}a_{2,0}.
\end{equation}
Here, recall that the vector $b^\mu,$ introduced in Eq. \eqref{eqn:Defb} as part of the most general solution to Eq. \eqref{eqn:MomCasCorr}, is entirely undetermined and independent of the deformation $d^\mu.$ 

Comparing with \cite{Wagner:2023fmb}, the effective Schr\"odinger equation for uncharged spinors differs from the nonrelativistic limit of the Klein-Gordon equation for these MDRs inasmuch as the inertial mass attains an additional deformation. This can only be avoided for $b_0=0$. However, the choice $b_0=0$ breaks Lorentz-invariance in a way which is unrelated to the deformation vector $d^\mu.$ In particular, it singles out the reference system with respect to which we take the nonrelativistic limit, as special in the universal context. Such a choice can only be justified if the underlying theory is invariant under a deformation of Poincar\'e symmetry (as in DSR) such that the deformed Lorentz transformation $\Lambda^{~\nu}_{\mu}=\Lambda^{(0)\nu}_{~~\mu}+\ell\Lambda^{(1)\nu}_{~~\mu}$ acts on $\hpi_\mu$ as 
\begin{equation}
    \Lambda^{~\nu}_{\mu}\hpi_\nu=\Lambda^{(0)\nu}_{~~\mu}\hk_\nu+\ell\left(\Lambda^{(1)\nu}_{~~\mu}\hk_\nu+\Lambda^{(0)\nu}_{~~\mu}\hpi^{(1)}_\nu\right)=\hk'_\mu+\ell\hpi^{(1)\prime}_\mu.
\end{equation}
Then, we could achieve $\hpi^{(1)\prime}_\mu\propto b'_\mu$ with $b'_0=0$ for all transformations $\Lambda$ with respect to all possible reference systems. 

\begin{acknowledgements}
F.I. and L. P. acknowledge support by MUR (Ministero dell'Universit\`a e della Ricerca) via the project PRIN 2017 ``Taming complexity via QUantum Strategies: a Hybrid Integrated Photonic approach'' (QUSHIP) Id. 2017SRNBRK. F. I. acknowledges support by MUR via the project PRIN PNRR 2022 ``Harnessing topological phases for quantum technologie'' Id. P202253RLY. P.B., F.W. and L.P. acknowledge networking support by the COST Action CA18108. 
\end{acknowledgements}

\bibliographystyle{utphys}
\bibliography{bib.bib}

\end{document}